\documentclass[showpacs,preprintnumbers,english,amsmath,amssymb]{revtex4}
\usepackage[T1]{fontenc}
\usepackage[latin1]{inputenc}
\usepackage{graphicx}
\usepackage{amssymb}
\usepackage{verbatim}
\usepackage{epic,eepic}
\usepackage{babel}
\usepackage{color}
\usepackage{rotating}

\def\dis{distribution}
\def\pt{p_T}

\begin{document}

\def\Journal#1#2#3#4{{#1}, {#2}, {#3}, {#4}.}
\def\ADEP{Advances in High Energy Physics}
\def\ANP{Adv. Nucl. Phys.}
\def\ARNPS{Ann. Rev. Nucl. Part. Sci.}
\def\CTP{Commun. Theor. Phys.}
\def\CPL{Chin. Phys. Lett.}
\def\EPJA{The European Physical Journal A}
\def\EPJC{The European Physical Journal C}
\def\IJMPA{International Journal of Modern Physics A}
\def\IJMPE{International Journal of Modern Physics E}
\def\JCHP{J. Chem. Phys.}
\def\JCP{Journal of Computational Physics}
\def\JHEP{Journal of High Energy Physics}
\def\JPCS{Journal of Physics: Conference Series}
\def\JPG{Journal of Physics G: Nuclear and Particle Physics}
\def\NATURE{Nature}
\def\NC{La Rivista del Nuovo Cimento}
\def\NCA{IL Nuovo Cimento A}
\def\NPA{Nuclear Physics A}
\def\NPB{Nuclear Physics B}
\def\NST{Nuclear Science and Techniques}
\def\PA{Physica A}
\def\PAN{Physics of Atomic Nuclei}
\def\PHY{Physics}
\def\PRA{Phys. Rev. A}
\def\PRC{Physical Review C}
\def\PRD{Physical Review D}
\def\PLA{Phys. Lett. A}
\def\PLB{Physics Letters B}
\def\PLD{Phys. Lett. D}
\def\PRL{Physical Review Letters}
\def\PL{Phys. Lett.}
\def\PREV{Phys. Rev.}
\def\PREP{\em Physics Reports}
\def\PROG{Progress in Particle and Nuclear Physics}
\def\RPP{Rep. Prog. Phys.}
\def\RDNC{Rivista del Nuovo Cimento}
\def\RMP{Rev. Mod. Phys}
\def\SCIENCE{Science}
\def\ZPA{Z. Phys. A.}

\def\ANN{Ann. Rev. Nucl. Part. Sci.}
\def\ANNAST{Ann. Rev. Astron. Astrophys.}
\def\AP{Ann. Phys}
\def\APJ{Astrophysical Journal}
\def\APJS{Astrophys. J. Suppl. Ser.}
\def\EJP{Eur. J. Phys.}
\def\LANC{Lettere Al Nuovo Cimento}
\def\NCA{Nuovo Cimento A}
\def\PHYS{Physica}
\def\NP{Nucl. Phys}
\def\MATH{J. Math. Phys.}
\def\JPAM{J. Phys. A: Math. Gen.}
\def\PRO{Prog. Theor. Phys.}

\title{The effect of minijet on hadron spectra and azimuthal anisotropy in heavy-ion collisions}

\author{Lilin Zhu\footnote{Email address: zhulilin@scu.edu.cn}}
\affiliation{
College of Physical Science and Technology, Sichuan University, Chengdu 610064, People's Republic of China.}


\begin{abstract}
Here I review the transverse momentum distributions of identified hadrons produced in Au-Au collisions at RHIC and Pb-Pb collisions at LHC in the framework of recombination model. Minijets play an important role in generating shower partons in the intermediate $p_T$ region. At LHC, the resultant soft shower partons are even found to dominate over the thermal partons in the non-strange sector. The azimuthal anisotropy of the produced hadrons could also be explained as the consequence of the effects of minijets. Harmonic analysis of the $\phi$ dependence leads to $v_n(p_T, b)$ that can be well produced without reference to flow.

\end{abstract}


\maketitle

\section{Introduction}
Theoretical investigation of hadron production in heavy-ion
collisions at high energies is usually separated into different
camps, characterized by the regions of transverse momenta $p_T$ of
the produced hadrons. At low $p_T$ statistical hadronization and
hydrodynamical models are generally used \cite{pbm, ph, kh, tat}. At high
$p_T$ jet production and parton fragmentation with suitable
consideration of medium effects in perturbative QCD are the
central themes \cite{pq1, pq2, ia, mt}. The two approaches have been studied
essentially independent of each other with credible success in
interpreting the experimental data. At the intermediate and lower $p_T$ region, recombination or coalescence (ReCo) model has been found to be more reasonable in heavy-ion collisions \cite{GKL2003, FMNB2003, HY2004}. The three ReCo models have some differences in details, but they are physically very similar. Refs. \cite{GKL2003, FMNB2003} gave the Wigner functions of produced hadrons with the phase-space distributions of the constituents at the freeze out. The recombination and fragmentation are treated as independent components of the hadronization. The ReCo models are also used to study light nuclei production in heavy-ion collisions \cite{Gyulassy:1982pe, Indra00, Zhang:2009ba, Zhu:2015voa}.  This review will follow the formulation of the recombination model (RM) by Hwa-Yang \cite{HY2004}. This model is one dimensional on the basis that non-collinear partons have very low probability of coalescence, and is simple enough to include fragmentation as a component of recombination of shower partons so that there is a smooth transition from low to high $p_T$. 

The aim of this review is to give an overview how to treat the hadornization for the whole $p_T$ region for Au-Au collisions at RHIC and Pb-Pb collisions at LHC with Hwa-Yang recombination model. In Pb-Pb collisions, there are much more soft partons and hard jets produced. Therefore, the hadronization at LHC is drastically different from that at RHIC. In the recombination model, there are two types of partons: thermal (T) and shower (S). The minijets generate shower partons after emerging from the medium surface. Those shower partons recombine with themselves or with the thermal partons in various combination to form hadron. Therefore, the effect of minijets on hadron production can't be ignored.  Beside the $p_T$ distribution, the azimuthal anisotropy have been studied by considering the ridges by semihard scattering in Ref. \cite{CHY2008}. Quark number scaling of $v_2$ was also found to be only approximately valid at low $p_T$, but was broken at intermediate $p_T$. With more careful consideration of momentum degradation on semihard parton, we will review the effect of minijets on azimuthal dependence in non-central Au-Au collisions. Conventional description of azimuthal anisotropy doesn't consider the effects of jets. In this review the details of calculations will be not shown, and adequate referencing is provided to guide the interested reader to the original papers where details can be found.

\section{The recombination model}
First, we show a brief summary of the main equations of the recombination model, which are collected in Refs. \cite{HY2004, HZ2011, HZ2012, ZH2013}. The invariant $p_T$ distributions of meson and baryon at midrapidity are
\begin{eqnarray}
p^0{dN^M\over dp_T}&=&\int {dp_1\over p_1}{dp_2\over p_2} F_{q_1\bar q_2}(p_1,p_2) R_{q_1\bar q_2}^M(p_1,p_2,p_T) \label{21} \\
p^0{dN^B\over dp_T}&=&\int \left[\prod_{i=1}^3 {dp_i\over p_i} \right] F_{q_1q_2q_3}(p_1,p_2,p_3) {R}_{q_1q_2q_3}^B(p_1,p_2,p_3,p_T)    \label{22}
\end{eqnarray}
where $p_i$ is the transverse momentum of one of the coalescing partons. $R^M $ and $R^B$ are the recombination functions of meson and baryon, respectively \cite{HY2004, HZ2011}, which were introduced a long time ago and determined by the effects of dressing and hadronic structure \cite{DH1977}. The LHS of Eqs.\ (\ref{21}) and (\ref{22}) are the invariant $p_T$ distributions of meson and baryon, respectively, averaged over $\eta$ at midrapidity. The $\phi$ dependence has been averaged over, so $dN^h/p_Tdp_T$ should be identified with the experimental $dN/2\pi p_Tdp_T$. 

The central issue in the formalism is the determination of the parton distributions $F_{q_1\bar{q}_2}$ and $F_{q_1q_2q_3}$ just before hadronization.  $\cal T$ and $\cal S$ to denote the thermal and shower partons invariant distributions at the late time just before hadronization, respectively. The thermal parton contains the medium effect, while the shower parton is due to the semihard and hard scattered partons. The  shower partons we consider are the fragmentation products of the hard and semihard partons that emerge from the surface after momentum degradation. They are distinguished from the thermal partons  that are  in their environment. Taking into account the recombination of different types of partons, we have
\begin{eqnarray}
F_{q_1\bar q_2}&=&{\cal TT+TS+SS}  \label{23} \\
F_{q_1q_2q_3}&=&{\cal TTT+TTS+TSS+SSS}    \label{24}
\end{eqnarray}
The two shower partons in $\cal SS$ and $\cal TSS$ are probably from one or two jets. The three shower partons in $\cal SSS$ could be even from three different jets. Here we will not consider three jets contribution to the hadron production.

The thermal parton distribution is
\begin{eqnarray}
{\cal T}(p_1) = p_1{ dN^T_q\over dp_1} = Cp_1e^{-p_1/T}    \label{25}
\end{eqnarray}
where $C$ has the dimension of inverse momentum and $T$ is the inverse slope parameter that should not be treated as the same as the conventional temperature in hydrodynamical model. They could be fixed by the experimental data at low $p_T$. The dimensionless prefactor $Cp_1$ is necessary to yield pure exponential behavior for the hadron distribution $dN^{h}/p_Tdp_T$. On the other hand, the properties of shower parton distribution depend on the collisions energy, not only due to many more hard and semihard partons produced but also the quenching effect of hot dense medium.

The shower parton distribution after integration over jet momentum $	q$ and summed over all jets is 
\begin{eqnarray}
{\cal S}^j(p_2)=\int {dq\over q}\sum_i \hat F_i(q) S_i^j(p_2/q),  \label{26}
\end{eqnarray}
 where $\hat F_i(q)$ is the distribution of hard or semihard parton of type $i$ at the medium surface after momentum degradation while transversing the medium but before fragmentation. It also depends on the centrality and medium property, which will be discussed in the next section. $\hat F_i(q)$ was introduced previously for collisions at RHIC for any centrality \cite{HZ2012}, but also modified to suit the description of the physics at LHC \cite{ZH2014}. $\hat F_i(q, b)$ is defined as the average of $\bar F_i(q, \phi, b)$ over $\phi$
\begin{eqnarray}
\hat F_i(q, b)=\frac{1}{2\pi}\int_0^{2\pi}d\phi\bar F_i(q, \phi, b), \label{48}
\end{eqnarray}
 The average parton distribution $\bar F_i(q, \phi, b)$ will be discussed in Sec. III. $S_i^j(z)$ is the shower-parton distribution (SPD) in a jet of type $i$ fragmentation into a parton of type $j$ with momentum fraction $z$. SPD is determined by the fragmentation function (FF) on the basis that hadrons in a jet are formed by recombination of the shower partons in the jet \cite{HY20042}. In particular, the recombination of a quark $j$ with an antiquark $\bar j$ in a jet of type $i$ forms a pion, for which the FF is $D_i^{\pi}(z_j+z_{\bar j})$. The numerical form for $S_i^j(z)$ can therefore be calculated from the data on $D_i^{\pi}$ and the RF of pion:
 \begin{eqnarray}
 xD_i^{\pi}(x)=\int\frac{dx_1}{x_1}\frac{dx_2}{x_2}\large\{S_i^q(x_1), S_i^{\bar q}(\frac{x_2}{1-x_1})\large\}R^{\pi}(x_1, x_2, x).
 \end{eqnarray}
where the curly brackets denote symmetrization of the leading parton momentum fractions $x_1$ and $x_2$. Ref. \cite{HY20042} gave the parametrization of $S_i^j$ for all partons. Once the SPDs are known, one can consider the possibility that a shower parton recombines with a thermal parton and thus give a more complete description of hadronization at intermediate $p_T$ region. 

Only considering the $\cal{TT}$ ($\cal{TTT}$) component for pion (proton), we have been able to fit the pion and proton spectra for $1<p_T<2$ GeV/c in Au-Au collisions at 200 GeV with a common value of the inverse slope in Eq.\ (\ref{25}) \cite{HZ2012}. For $p_T<1$ GeV/c there is resonance contribution which couldn't be calculated in RM, while for $p_T>2$ GeV/c shower parton contributions invalidate the approximation of $F_{q\bar q}$ and $F_{uud}$ by $\cal {TT}$ and $\cal{TTT}$, respectively. As we shall see below, the situation of dominance by $\rm TT$ and $\rm TTT$ recombination changes when the collision energy is increased tenfold, whereby $\rm TS$ and $\rm TTS$ can no longer be neglected even for the low $p_T$ region at LHC. But before discussing the SPD, we need to have a clear picture for the process of momentum degradation of a semihard parton when transferring the hot dense medium.

\section{the momentum degradation}
Refs. \cite{ZH2013, ZH2014} gave a detailed explanation of the momentum degradation on a semihard or hard parton for any centrality and at any angle $\phi$. Here we show a brief summary on it. The average parton distribution $\bar F_i(q, \xi, b)$ for the momentum $q$ at the medium surface is defined as,
\begin{eqnarray}
\bar F_i(q,\phi,b) =\int d\xi P(\xi,\phi,b)F_i(q,\xi)= \int d\xi P(\xi,\phi,b)\int dk kf_i(k) G(k,q,\xi) ,     \label{31}
\end{eqnarray}
which averages over all $\xi$ with the weighting function $P(\xi, \phi, b)$ being the probability of having $\xi$ at $\phi$ and $b$. The dynamical length $\xi$ carries all the information on geometry and dynamics through $P(\xi, \phi, b)$. This probability $P(\xi, \phi, b)$ has been described in Refs. \cite{CHY2008, ZH2013}. 
$f_i(k)$ in the parton distribution $F(q, \xi)$ is the parton density in the phase space $kdk$ at the point of creation, $k$ being the initial momentum of the hard or semhard parton $i$. $G(k,q,\xi)$ is the momentum degradation function from  the initial parton momentum $k$ to the final momentum $q$ at the medium surface, 
\begin{eqnarray}
G(k,q,\xi) = q\delta(q-ke^{-\xi}).     \label{33}
\end{eqnarray}

The initial momentum distributions have been determined in Ref.\ \cite{sgf2003} for Au-Au collisions at 200 GeV and Pb-Pb collisions at 5.5 TeV. They are parametrized in the form
\begin{eqnarray}
f_i(k)=K\frac{C}{(1+k/B)^{\beta}}. \label{34}
\end{eqnarray}
To obtain $\ln A$, $B$ and $\beta$ for Pb-Pb collisions at $\sqrt{S_{NN}}=2.76$ TeV, we made logarithmic interpolations of the parameters between the two energies. The parameters for Au-Au collisions at $\sqrt{S_{NN}}=200$ GeV and Pb-Pb collisions at $\sqrt{S_{NN}}=2.76$ TeV are shown in Table I with $K=2.5$.

\begin{table}
\tabcolsep0.2in
\begin{tabular}{|c|c|c|c|c|}
\hline
 & & $C$ [1/GeV$^2$] &  $B$ [GeV]  &$\beta$\\ 
 \hline
 & $u$ &9.113$\times$10$^2$ &1.459 &7.679 \\
& $d$ &9.596$\times$10$^2$ &1.467 &7.662\\
Au-Au& $s$&1.038$\times$10$^2$ &1.868 &8.642 \\
& $\bar u$&2.031$\times$10$^2$ &1.767 &8.546\\
& $\bar d$&2.013$\times$10$^2$ &1.759 &8.566 \\
& $g$&4.455$\times$10$^3$ &1.7694 &8.610 \\
 \hline
 & $u$ &1.138$\times$10$^4$ &0.687 &5.67 \\
& $d$ &1.266$\times$10$^4$ &0.677 &5.66\\
Pb-Pb& $s$&0.093$\times$10$^4$ &1.05 &6.12 \\
& $\bar u$&0.24$\times$10$^4$ &0.87 &5.97 \\
& $\bar d$&0.23$\times$10$^4$ &0.88 &5.99 \\
& $g$&6.2$\times$10$^4$ &0.98 &6.22 \\
 \hline
 \end{tabular}
 \caption{Parameters for the minijet distribution $f_i(k)$ in Eq.\ (\ref{34}) at $y=0$ for Au-Au collisions at 200 GeV \cite{sgf2003} and Pb-Pb collisions at 2.76 TeV \cite{ZH2014}.} \label{table1}
 \end{table}

The connection between geometry and dynamics is imbedded in the probability function $P_i(\xi,\phi,b)$. 
The geometrical path length $\ell$ is 
\begin{eqnarray}
\ell(x_0, y_0, \phi, b)=\int_0^{t_1(x_1, y_1)}dt D(x(t), y(t)). \label{35}
\end{eqnarray}
It is calculable from nucleon geometry. The transverse coordinate $(x_0, y_0)$ is the initial point of creation of a hard parton, and $(x_1, y_1)$ is the exit point. The integration is weighted by the local density, $D(x, y)$, along the trajectory, which is marked by the variable $t$ that does not denote time. As the medium expands, the end point $t_1(x_1, y_1)$ increases, but $D(x(t), y(t))$ decreases, so $\ell$ is insensitive to the details of expansion dynamics. The dynamical path length $\xi$ is proportional to $\ell$, but is to be averaged over all initial points $(x_0, y_0)$, 
\begin{eqnarray}
P_i(\xi, \phi, b)=\int dx_0dy_0Q(x_0, y_0, b)\delta(\xi-\gamma_i\ell(x_0, y_0, \phi, b)) \label{36}
\end{eqnarray}
where $Q(x_0, y_0, b)$ is the probability that a hard (or semihard) parton is produced at $(x_0, y_0)$, calculable from nucleon thickness functions \cite{ZH2013}. The only parameter that we cannot calculate is $\gamma_i$, which incorporate the effects of energy loss during the passage of the parton through the non-uniform and expanding medium. The average dynamical path length $\bar\xi_i$, defined by
\begin{eqnarray}
\bar\xi_i(\phi, b)=\int d\xi\xi P(\xi, \phi, b), \label{37}
\end{eqnarray}
depends on geometry, and is proportional to $\gamma_i$. Thus, $\hat F_i(q, b)$ can be calculated once $\gamma_i$ are specified.

In treating hadron production at RHIC we chose suitable values of $\gamma_i$ for gluon and quark, and obtained excellent fits of the $p_T$ distributions of $\pi, K, p$ for $p_T<10$ GeV/c at six centralities \cite{ZH2013}. $\gamma_g=0.14$ for gluon and $\gamma_q=0.07$ for all light quarks. Because $\bar\xi_i(\phi, b)\propto\gamma_i$, we have $\bar\xi_g(\phi, b)/\bar\xi_q(\phi, b)=2$, which directly implies that gluons on average lose the same fraction of momentum as quarks do in half  the distance of traversal through the nucleon medium. That turned out to be an important factor in enabling us to reproduce both the pion and proton spectra because at intermediate $p_T$ pions are more affected by semihard gluon minijets, while protons are more by quark minijets \cite{HZ2012}. 

To extend the treatment of momentum degradation to collisions at LHC, $\gamma_i$ couldn't be expected to be the same as at RHIC any more. The data at LHC \cite{ALICE20132} suggest that jet quenching becomes less severe at higher momentum, so $\gamma_i$ should decrease as the hard parton momentum increases. Hence, we parametrize $\gamma_g$ as
\begin{eqnarray}
\gamma_g(q)=\frac{\gamma_0}{1+q/q_0}, \label{38}
\end{eqnarray}
We continue to set $\gamma_q=\gamma_g/2$ as at RHIC. $\gamma_0=0.8$ and $q_0=10$ GeV/c are determined by the experimental data of $\pi$ distribution at 0-5\% in Pb-Pb collisions. The two parameters were also used to describe the production of other hadrons, such as $K$, $p$, $\Lambda$, $\phi$, $\Xi$ and $\Omega$. The fits were very excellent. We also use Eq. (\ref{38}) to reconsider $\pi$ and $p$ distributions in Au-Au collisions at 200 GeV for 0-10\% centrality with $\gamma_0=0.6$ and $q_0=10$ GeV/c. 

\begin{figure*}
        \centering
        \begin{tabular}{ccc}
        \includegraphics[width=0.45\textwidth]{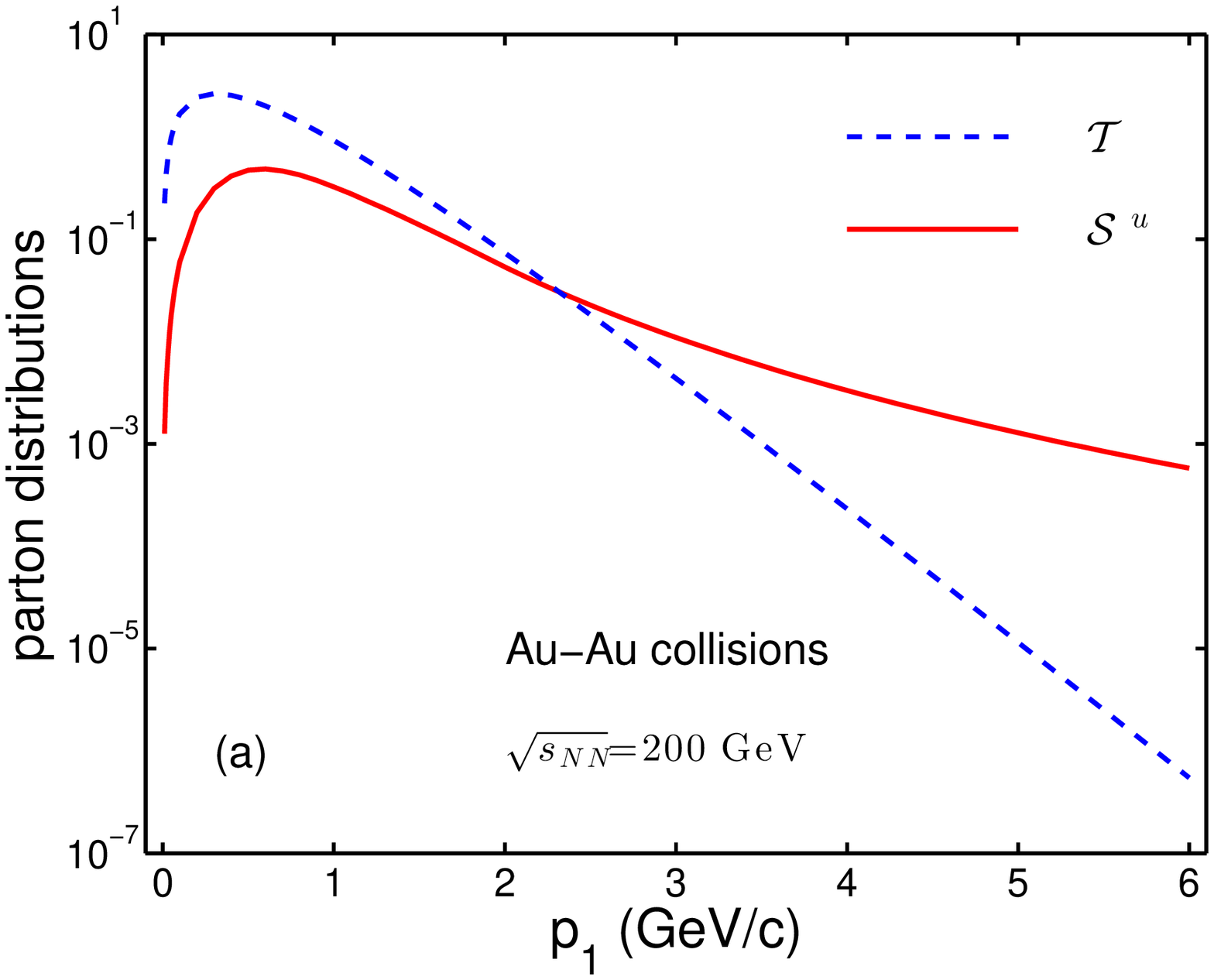}
          \includegraphics[width=0.45\textwidth]{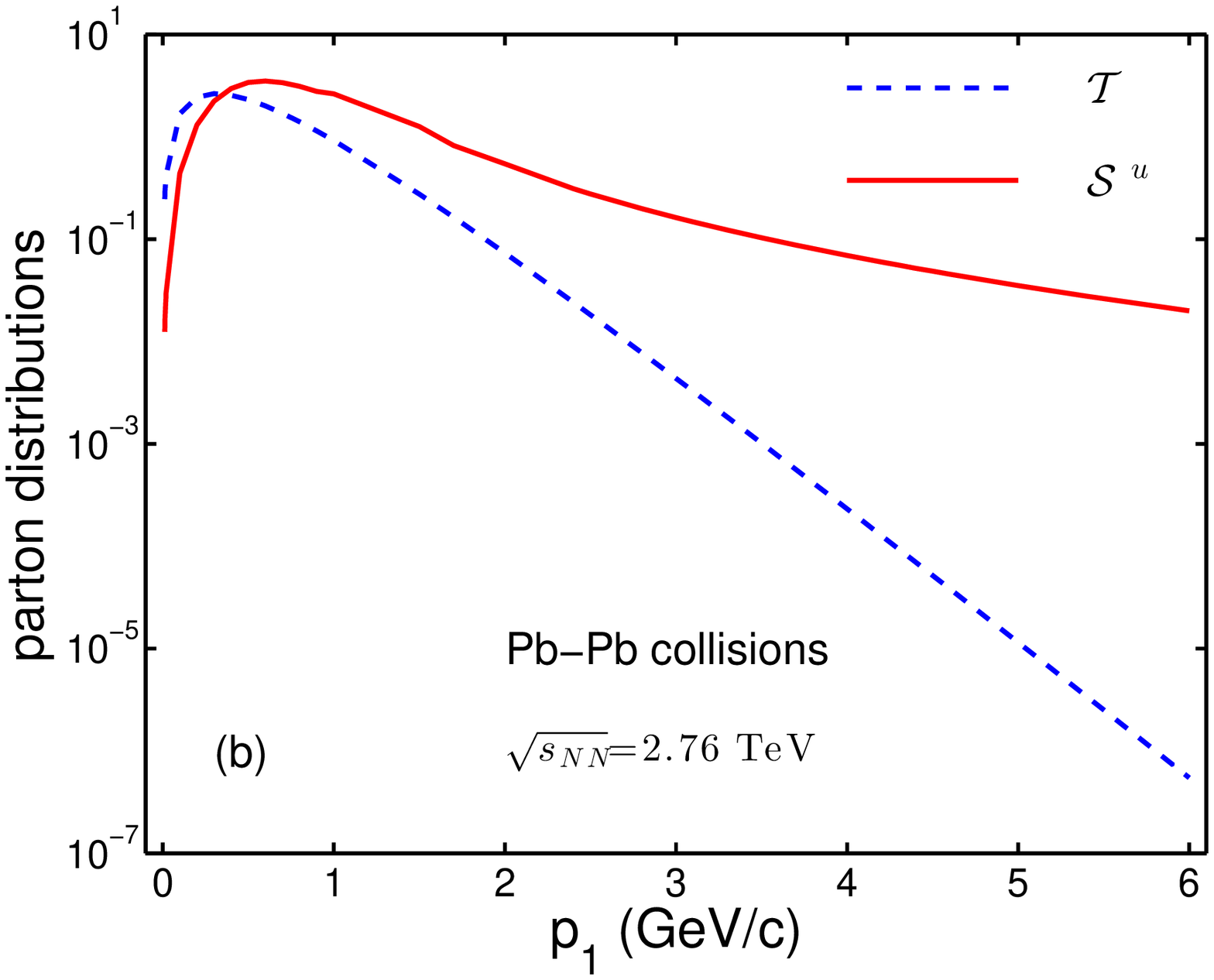}
               \end{tabular}
\caption{(Color online) The thermal distribution $\mathcal{T}(p_1)$ and shower parton distribution $\mathcal{S}^u$ are shown for Au-Au collisions at RHIC (a) and Pb-Pb collisions at LHC (b).}\label{SPD}
    \end{figure*}

 After determine the momentum degradation of minijets, we could obtain the invariant shower-parton distribution $\mathcal{S}^j$ in Eq. (\ref{26}) after intergrating over $q$ and summing over all initiating partons $i$.  As shown in Fig. \ref{SPD}(b), $\mathcal S(p_1)$ dominates over $\mathcal T(p_1)$ for all $p_1>0.5$ GeV/c at LHC, while at RHIC the cross over does not occur until $p_1>2$ GeV/c in Fig. \ref{SPD}(a). It means the shower partons play an important role even at low $p_T$ region at LHC. This is the most remarkable  feature about the parton distribution at LHC. The dominance of $\mathcal S(p_1)$ is so important that it reorients our thinking about hadron production at low and intermediate $p_T$. In essence, minijets are so copiously produced at LHC that their effects at low $p_T$ cannot be ignored, thus posing a substantive question on the meaningfulness of any hydrodynamical study without taking minijets into account.
 
\section{Hadronic Spectra}
The momentum degradation discussed in the last section was applied not just to Au-Au collisions at RHIC, but also to Pb-Pb collisions at LHC. The transverse momentum distributions at LHC were investigated in Ref. \cite{HZ2011}, but it was for a limited range of $p_T<5$ GeV/c and was based on a simple assumption about the momentum degradation, which was not reasonable for high $p_T$ region.  The formalism for recombination of thermal and shower partons in the two colliding systems are the same. Refs. \cite{HZ2011, ZH2014} only considered the central collision at Pb-Pb collisions, while for Au-Au collisions we have generalized to the non-central collisions \cite{ZH2013}. 

The thermal parton distribution is shown in Eq. (\ref{25}). The inverse slope $T$ is independent of the centrality, since $\mathcal{T}$ is the thermal parton distribution at the time of hadronization and has the same momentum dependence at any centrality. Furthermore, the thermal partons include the soft partons generated by hard and semihard partons as they transverse the medium and have thermalized with the bulk partons by the end of the deconfined phase. When those thermal partons are dilute enough and be ready for confinement through recombination, their local properties are no longer sensitive to the colliding system. Hence, we use the same form of thermal parton distribution $\mathcal{T}$ for RHIC and LHC. The values of $C$ and $T$ in the thermal parton distribution are used for calculating the spectra for $p_T>1$ GeV/c. At lower $p_T$ the pion distribution is lower than the data, which is undoubtedly related to the extra low traverse momentum partons created at LHC that we cannot easily include in our parametrization. The normalization factor $C$ is dependent on the centrality. At LHC we use the same centrality dependence for $C$ as at RHIC \cite{ZH2013, ZH2014},
\begin{eqnarray}
C(N_{part})=3.43N_{part}^{0.32}.
\end{eqnarray}
Here we only review the $p_T$ distributions for $\pi$ and $p$ at RHIC and LHC. The results for other mesons and baryons could be found in Ref. \cite{HZ2011, ZH2013, ZH2014}. It should emphasis that the above equations for hadronization in recombination model are applied to the final stage of the evolution of the colliding system when the hot dense medium is very low. The hadrons are formed by the recombination of quarks and antiquarks, and all gluons are converted to quark-antiquark pairs, so there is no gluon at the final stage. 

\subsection{pion}
\begin{eqnarray}
{dN^{TT}_{\pi}\over p_Tdp_T} &=&\frac{C^2}{6}e^{-p_T/T} ,   \label{41}
\end{eqnarray}
\begin{eqnarray}
{dN_{\pi}^{TS}\over p_Tdp_T} &=& {C\over p_T^3} \int_0^{p_T} dp_1 p_1e^{-p_1/T}
\left[{\cal S}^{u}(p_T-p_1) +{\cal S}^{\bar d}(p_T-p_1)\right] ,    \label{42} \\
{dN^{{SS}^{1j}}_{\pi}\over p_Tdp_T} &=& {1\over p_T} \int {dq\over q^2} \sum_i \hat{F}_i(q)D^{\pi}_i(p_T,q) ,   \label{43}\\
{dN_{\pi}^{{SS}^{2j}}\over p_Tdp_T} &=& {\Gamma\over p_T^3} \int_0^{\pt} dp_1  {\cal S}^{u}(p_1) {\cal S}^{\bar d}(p_T-p_1) ,    \label{44} 
\end{eqnarray}
where $\Gamma$ is the probability that two shower partons can recombine.

\subsection{proton production} 
\begin{eqnarray}
\frac{dN_p^{TTT}}{p_Tdp_T}=g_{st}^pg_pB(\alpha+2, \beta+2)B(\alpha+2, \alpha+\beta+4)\frac{C^3p_T^2}{m_T^p}e^{-p_T/T}, \label{45}
\end{eqnarray}
with $g_{st}=1/6$, $\alpha=1.75$, $\beta=1.05$ and $g_p=[B(\alpha+1, \alpha+\beta+2)B(\alpha+1, \beta+1)]^{-1}$. $B(a, b)$ is the Beta function.
\begin{eqnarray}
{dN_p^{TTS}\over \pt d\pt}&=&{g_{st}^pg_p C^2\over m_T^p \pt^{2\alpha+\beta+3}} \int_0^{\pt} dp_1 \int_0^{\pt-p_1} dp_2\ e^{-(p_1+p_2)/T}  \nonumber \\
	&& \hspace{1cm} \times\left\{ (p_1p_2)^{\alpha+1}(\pt-p_1-p_2)^{\beta} {\cal S}^d(\pt-p_1-p_2)\right.\nonumber\\
&&\hspace{1cm}\left.+p_1^{\alpha+1}p_2^{\beta+1}(\pt-p_1-p_2)^{\alpha} {\cal S}^u(\pt-p_1-p_2)\right\}, \label{46}
\end{eqnarray}

\begin{eqnarray}
{dN_p^{{TSS}^{1j}}\over \pt d\pt}&=&{g_{st}^pg_p C\over m_T^p \pt^{2\alpha+\beta+3}} \int_0^{\pt} dp_1 \int_0^{\pt-p_1} dp_2\ e^{-p_1/T}  \nonumber \\
	&& \hspace{1cm} \times\left\{ p_1^{\beta+1}p_2^{\alpha}(\pt-p_1-p_2)^{\alpha} {\cal S}^{uu}(p_2,\pt-p_1-p_2)\right.\nonumber\\
&&\hspace{1cm}\left.+p_1(p_1p_2)^{\alpha}(\pt-p_1-p_2)^{\beta} {\cal S}^{ud}(p_2,\pt-p_1-p_2)\right\},\label{47}
\end{eqnarray}

with
\begin{eqnarray}
\mathcal{S}^{qq}(p_2, p_3)=\int\frac{dq}{q}\sum\limits_i\hat F_i(q){\rm S}_i^q(p_2, q){\rm S}_i^q(p_3, q-p_2).\label{48}
\end{eqnarray}

\begin{eqnarray}
{dN_p^{{SSS}^{1j}}\over \pt d\pt}=\frac{1}{m_p^T}\int\frac{dq}{q^2}\sum\limits_i\hat F_i(q)D_i^p(p_T, q),\label{49}
\end{eqnarray}

\begin{eqnarray}
{dN_p^{{TSS}^{2j}}\over \pt d\pt}&=&{g_{st}^pg_p  C\Gamma\over m_T^p \pt^{2\alpha+\beta+3}} \int_0^{\pt} dp_1 \int_0^{\pt-p_1} dp_2\ e^{-p_1/T}  \nonumber \\
	&& \hspace{1cm} \times\left\{ p_1^{\beta+1}p_2^{\alpha}(\pt-p_1-p_2)^{\alpha} {\cal S}^u(p_2) {\cal S}^{u}(\pt-p_1-p_2)\right.\nonumber\\
&&\hspace{1cm}\left.+p_1(p_1p_2)^{\alpha}(\pt-p_1-p_2)^{\beta} {\cal S}^u(p_2) {\cal S}^{d}(\pt-p_1-p_2)\right\}, \label{410}
\end{eqnarray}

\begin{eqnarray}
{dN_p^{{SSS}^{2j}}\over \pt d\pt}&=&{g_{st}^pg_p\Gamma \over m_T^p \pt^{2\alpha+\beta+3}} \int_0^{\pt} dp_1 \int_0^{\pt-p_1} dp_2  \nonumber \\
	&& \hspace{1cm} \times\left\{ p_1^{\beta}p_2^{\alpha}(\pt-p_1-p_2)^{\alpha} {\cal S}^d(p_1) {\cal S}^{uu}(p_2,\pt-p_1-p_2)\right.\nonumber\\
&&\hspace{1cm}\left.+(p_1p_2)^{\alpha}(\pt-p_1-p_2)^{\beta} {\cal S}^u(p_1) {\cal S}^{ud}(p_2,\pt-p_1-p_2)\right\}. \label{411}
\end{eqnarray}

\begin{figure*}
        \centering
        \begin{tabular}{ccc}
        \includegraphics[width=0.45\textwidth]{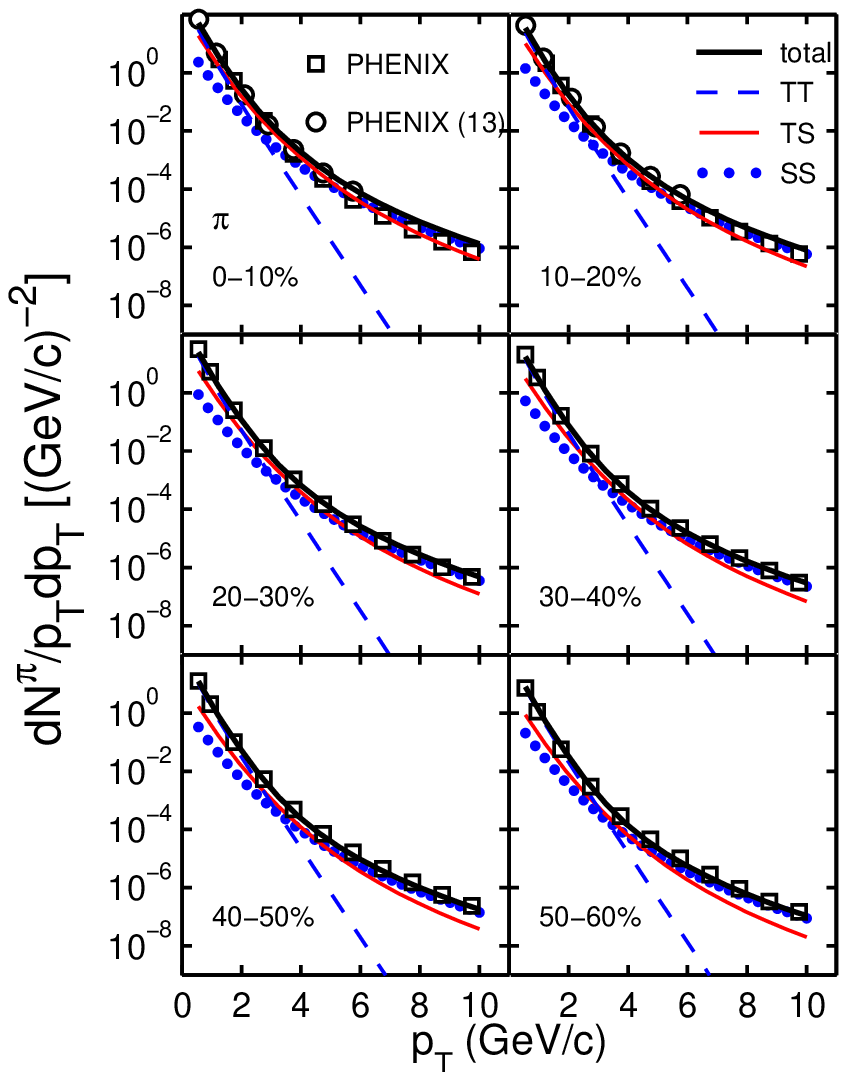}
          \includegraphics[width=0.45\textwidth]{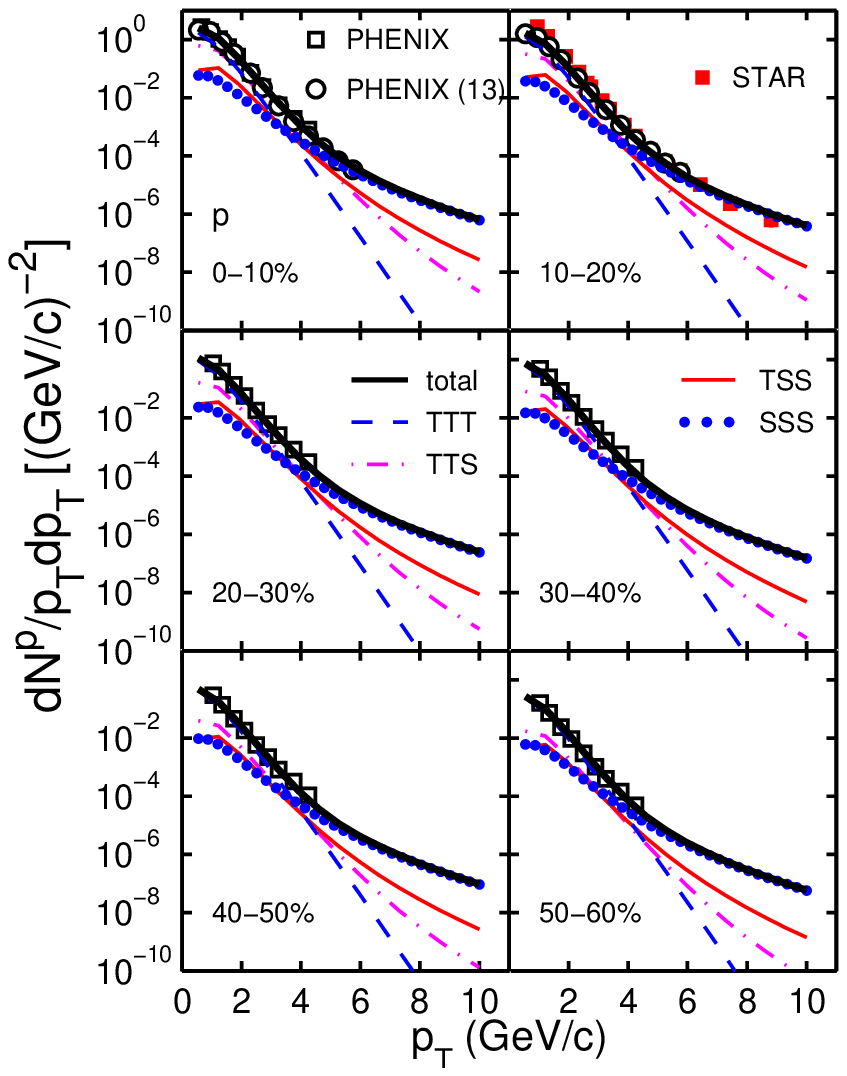}
               \end{tabular}
\caption{(Color online) Transverse momentum distributions for pion (left) and proton (right) for six centralities in Au-Au collisions at $\sqrt{S_{NN}}=200$ GeV. The data are from Refs. \cite{PHENIX2004, PHENIX2008, PHENIX2013, PHENIX2003, STAR2006}.}\label{spectra_RHIC_old}
    \end{figure*}

Fig. \ref{spectra_RHIC_old} shows the pion and proton spectra at Au-Au collisions at 200 GeV/c. It's amazing that we could fit the data for six centralities by only varying two parameters $C_0$ and $\omega$. In each case, TS, TTS and TSS components play crucial roles in uplifting the spectra in the intermediate $p_T$ region. For the high $p_T$ region, only one jet contribution is considered. The shower partons in SS and SSS are from one jet. The parameters $\gamma_g=0.14$ and $\gamma_q=0.07$ indicated more quark-type minijets survive the medium effect than the gluons. On the other hand, the difference of energy loss between quark and gluon minijets shows that the hadrons formed in recombination model are sensitive to the parton distributions. Pions depend more on gluons, while protons are dependent on quarks. The excellent fits for all centralities also show the success of recombination model. For Pb-Pb collision, we only have considered the central centrality 0-5\%. The parameters $\gamma_g$ and $\gamma_q$ are not constants any more. The two parameters $\gamma_0$ and $q_0$ in Eq. (\ref{38}) for the gluon degradation factor are crucial to get a good fit of pion and proton distributions at $p_T$ up to 20 GeV/c. That makes good sense in physics since the degradation of hard and semihard parton momenta is the central theme of heavy-ion physics at LHC. Ref. The minijets are so important when explaining the data at the whole $p_T$ region. Fig. 4 shows that pion and proton spectra could be well described with the choice of $\gamma_0=0.8$ and $q_0=10$ GeV/c. It's non-trivial to reproduce the data in such a wide range of $p_T$ and it's remarkable that the main input is just the momentum degradation factor $\gamma_g(q)$, which is not just good for pion and proton spectra, but also for all other hadrons, such as $K$, $\Lambda$, $\phi$, $\Xi$ and $\Omega$ \cite{ZH2014}. These results strongly supports the assertion that minijet production plays the dominate role in the structure of hadronic spectra. 

\begin{figure*}
        \centering
        \begin{tabular}{ccc}
        \includegraphics[width=0.45\textwidth]{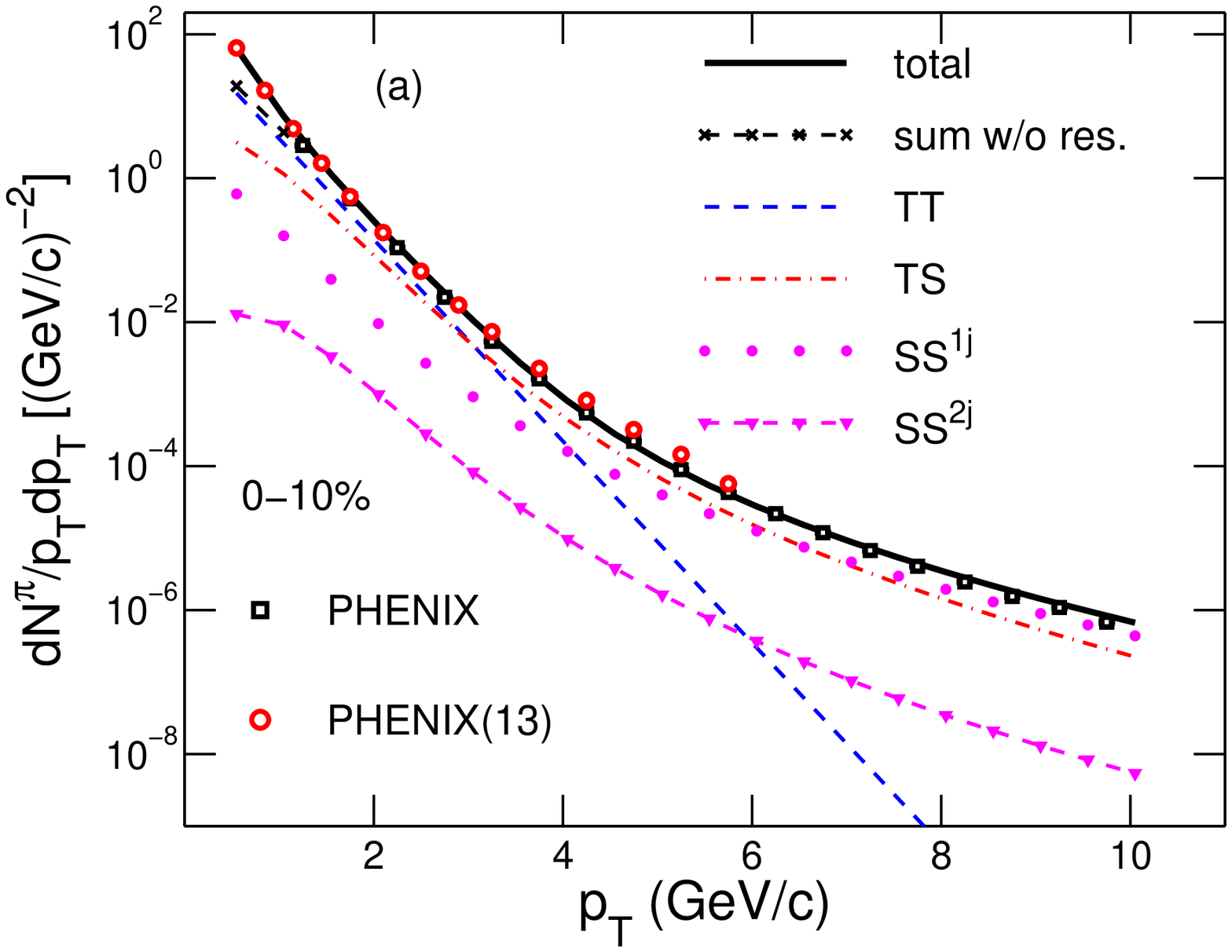}
          \includegraphics[width=0.45\textwidth]{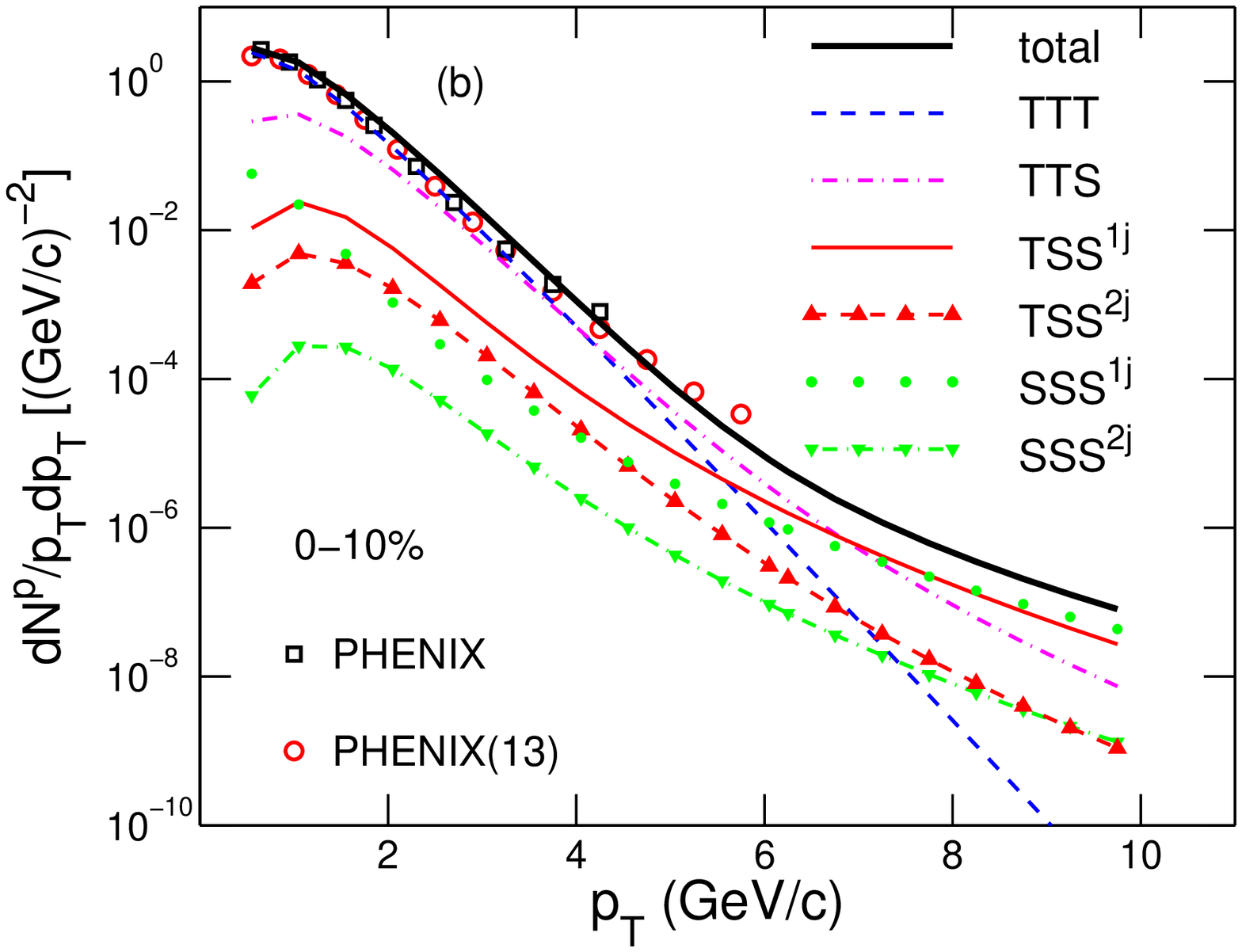}
               \end{tabular}
\caption{(Color online) Transverse momentum distributions for pion (a) and proton (b) for 0-10\% centrality in Au-Au collisions at $\sqrt{S_{NN}}=200$ GeV. The data are from Refs. \cite{PHENIX2004, PHENIX2008, PHENIX2013, PHENIX2003, STAR2006}.}\label{spectra_RHIC}
    \end{figure*}

\begin{figure*}
        \centering
        \begin{tabular}{ccc}
        \includegraphics[width=0.45\textwidth]{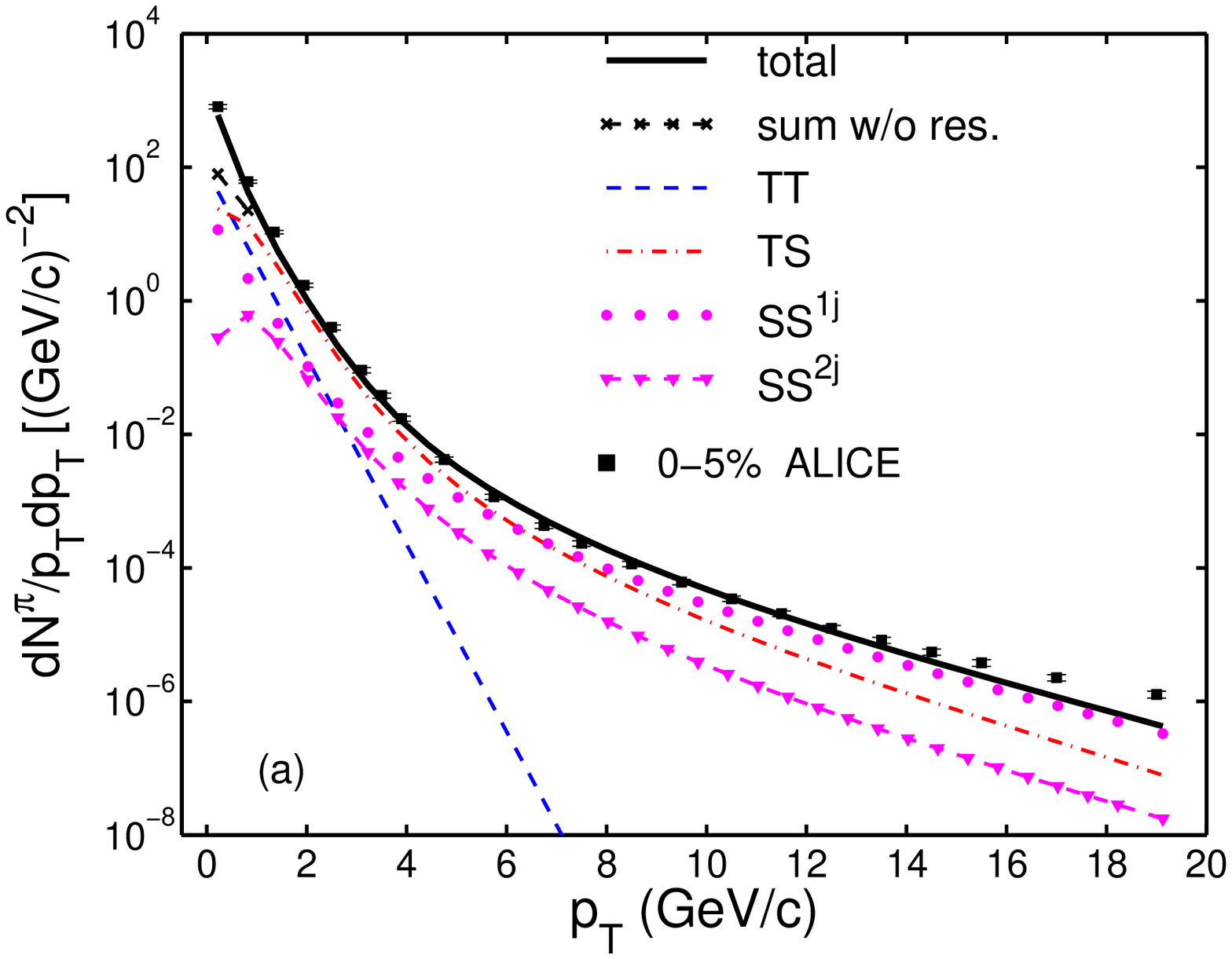}
          \includegraphics[width=0.45\textwidth]{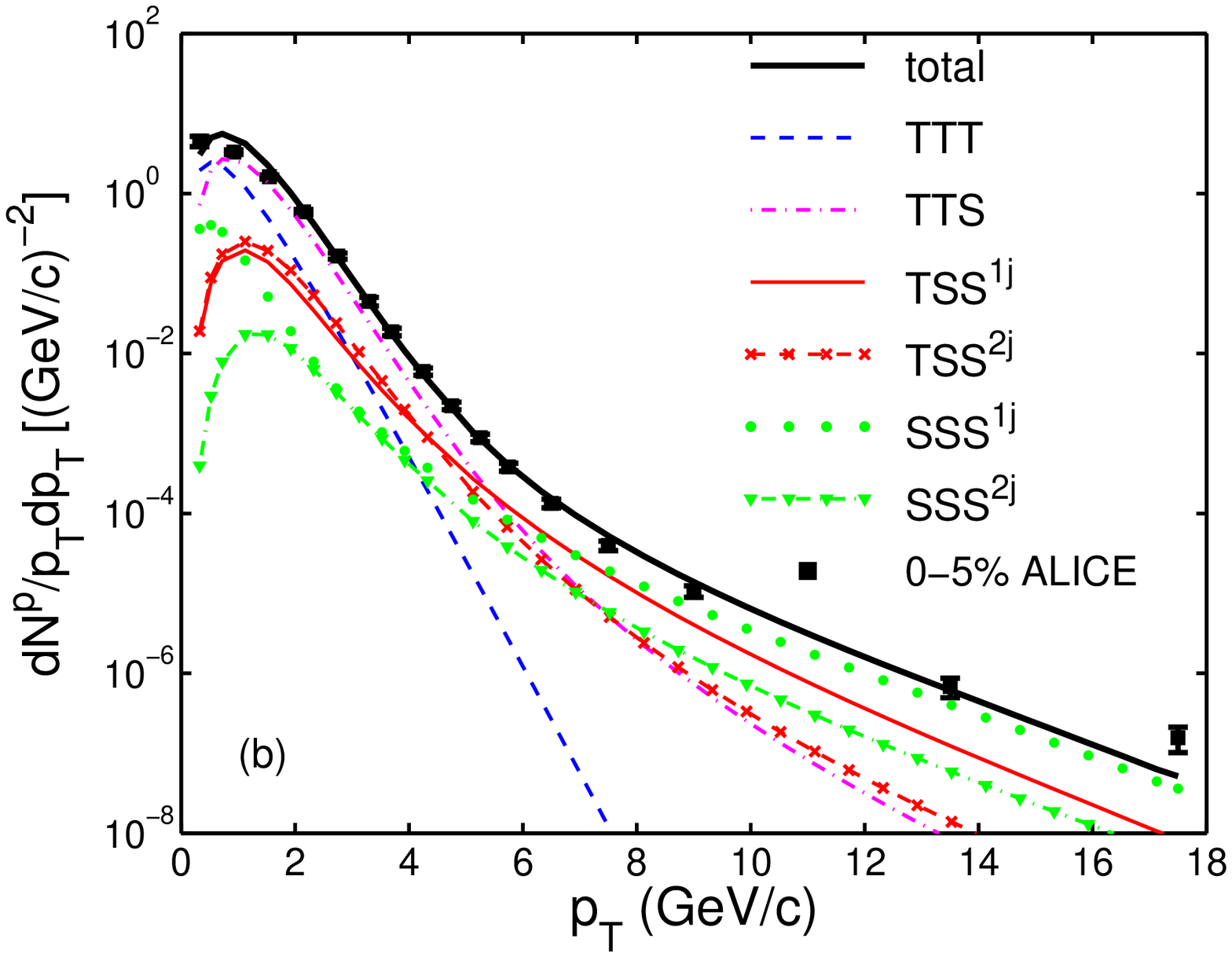}
               \end{tabular}
\caption{(Color online) Transverse momentum distributions for pion (a) and proton (b) for central centrality 0-5\% in Pb-Pb collisions at $\sqrt{S_{NN}}=2760$ GeV. The data are from Refs. \cite{ALICE2013}.}\label{Fig4}
    \end{figure*}

In order to make sensible comparison between LHC and RHIC results, we recalculated here the pion and proton distributions at RHIC, using the same description of the effects of energy loss on the shower partons, as shown Eq. (\ref{38}). The basic difference between Refs.\ \cite{ZH2013, ZH2014} is that $\gamma_g(q)$ is $q$ dependent as given in Eq.\ (\ref{38}). We also get good fit of the $\pi$ distribution in Au-Au collisions at 200 GeV for 0-10\% centrality, with $\gamma_0=0.6$ and $q_0=10$ GeV/c. Comparing Fig. \ref{spectra_RHIC}(a) to the pion distribution at LHC in Fig. \ref{Fig4}(a), one can see the drastic difference in $\rm TS$ relative to $\rm TT$ between the two cases. At RHIC $\rm TS$ crosses $\rm TT$ at $p_T\approx3$ GeV/c, whereas at LHC it occurs at $p_T\approx0.5$ GeV/c.
   
\section{azimuthal anisotropy}
The azimuthal dependence of single-particle distribution has been studied since the beginning of heavy-ion collisions \cite{STAR2005, PHENIX2004_KA}. For non-central collisions the almond-shaped average initial configuration leads to $\phi$ anisotropy. In a hydrodynamics picture the azimuthal anisotropy is understood in terms of pressure gradients \cite{SSD2005, HKH2002, TLHH2001, TN2004, HR2006}. Our approach showed that the azimuthal harmonics can be obtained by considering the azimuthal dependence of minijet and the related ridge effect. The results show the non-flow effects such as minijets are important. 

Let's use $\rho_h(p_T, \phi, b)$ to denote the single-particle distribution of hadron $h$ produced at midrapidity at impact parameter $b$,
\begin{eqnarray}
\rho^h(p_T,\phi,b) = {dN_h\over p_Tdp_Td\phi}(N_{\rm part}),     \label{51}
\end{eqnarray} 
Assuming the hadron distribution $\rho^h$ at low $p_T$ could be divided into three components:
\begin{eqnarray}
\rho^h(p_T,\phi,b) =B^h(p_T,b) + R^h(p_T,\phi,b) + M^h(p_T,\phi,b),     \label{52}
\end{eqnarray}
 referred to as base, ridge and minijet components, respectively.
$B^h(p_T,b)$ is azimuthally isotropic, which should not be confused with the $\phi$-dependent bulk distribution in the hydro description.
 $R^h(p_T,\phi,b)$ and $M^h(p_T,\phi,b)$ are  $\phi$ dependent.
 The first two components are due to the recombination of thermal partons (TT for pion and TTT for proton), while the third one is due to thermal-shower recombination (TS and TTS) \cite{HY2004}. $M^h(p_T,\phi,b)$ is dominant in the intermediate $\pt$ region ($2<\pt<6$ GeV/c), but is not negligible at low $p_T$. The normalized harmonic coefficients $v_n(p_T, b)$ can be calculated analytically, 
 \begin{eqnarray}
v_n^h(p_T,b) = \langle \cos n\phi \rangle_{\rho}^h = {\int_0^{2\pi} d\phi \cos n\phi\rho^h(p_T,\phi,b)\over \int_0^{2\pi} d\phi\rho^h(p_T,\phi,b)},     \label{53}
\end{eqnarray}
where $\rho^h(p_T, \phi, b)$ in our formalism has the three components given in Eq.\ (\ref{52}). We now describe the $\phi$ dependence of $R^h(p_T, \phi, b)$ and $M^h(p_T, \phi, b)$ separately.
 
 \subsection{second harnomic of $\phi$ anisotropy}
 $R^h(p_T, \phi, b)$ contains the $\phi$ anisotropy arising from the initial elliptical spatial configuration through the $S_2(\phi, b)$ function which transforms the spatial to momentum asymmetry. $S_2(\phi, b)$ is the segment of the surface through which the semihard parton can be emitted to contribute a ridge particle at $\phi$. More details on $S_2(\phi, b)$ could be found in \cite{HZ2010}. Since the elliptical axes need not coincide with the reaction plane that contains the impact parameter vector $\vec b$, we introduce a tilt angle $\psi_2$ and average over it. Then, we define $S(\phi, b)$ as

\begin{eqnarray}
S(\phi, b)&=&\tilde S_2(\phi, b)\left /{{1\over 2\pi} \int_0^{2\pi} d\phi \tilde S_2(\phi, b)}\right. \nonumber \\
&=&{2\over \pi}\int_{-\pi/4}^{\pi/4} d\psi_2 S_2(\phi-\psi_2, b)\left /{{1\over 2\pi} \int_0^{2\pi}{2\over \pi}\int_{-\pi/4}^{\pi/4} d\psi_2 S_2(\phi-\psi_2, b)}\right. \label{55}
\end{eqnarray}
 We now can write the ridge component of $\rho^h$ as
\begin{eqnarray}
R^h(p_T, \phi, b)=S(\phi, b) \bar R^h(\pt,b) ,  \label{56}
\end{eqnarray}
where $\bar R^h(\pt,b)$ is the second of two components of $dN_h^{\rm TT(T)}/\pt d\pt$. The exponential behavior of the first component, which is the $\phi$-independent base component $B^h(\pt,b)$, has  a lower $T_0$ than the overall $T$  for the sum of the two thermal terms described by Eq.\ (\ref{41}). Thus, the base thermal component is expressed as
\begin{eqnarray}
B^h(\pt,b) ={\cal N}_h(\pt,b)e^{-\pt/T_0},  \label{57}
\end{eqnarray}
the enhanced ridge component is
\begin{eqnarray}
\bar R^h(\pt,b)= {\cal N}_h(\pt,b)[e^{-\pt/T} -e^{-\pt/T_0}].  \label{58}
\end{eqnarray}
We emphasize that the only factor that  depends on the hadron type is $ {\cal N}_h(\pt,b)$, which represents the prefactor in Eqs. (\ref{41}) or (\ref{45}) before the exponential. It is a specific property of the recombination model that the exponential factors of the hadrons (whether $\pi$ or $p$) are inherited from those of the partons as discussed in the  preceding section. If we neglect the TS component for the sake of simplicity, since it is small at low $\pt$, we would have only the first two terms of $\rho^h(p_T, \phi, b)$ in Eq.\ (\ref{52}). In this case, we can obtain for $v_2^h(p_T, b)$ for hadron,
\begin{eqnarray}
v_2^h(p_T,b) ={\langle \cos2\phi \rangle_S\over Z^{-1}(p_T) + 1},     \label{59}
\end{eqnarray}
where
\begin{eqnarray}
\langle \cos2\phi \rangle_S &=& {1\over 2\pi} \int_0^{2\pi} d\phi \cos2\phi S(\phi,b),     \label{510} \\
Z(p_T) &=& {\bar R^h(\pt)\over B^h(\pt)}= e^{p_T/T'} - 1, \qquad T'={T_0 T\over T- T_0} .     \label{511}
\end{eqnarray}
These equations are remarkable, since the $b$ dependence resides entirely in Eq.\ (\ref{510}) and the $p_T$ dependence entirely in Eq.\ (\ref{511}). Furthermore, there is no explicit dependence on the hadron type.

 Fig. 5 showed the fits for pion and proton for the centrality 0-5\% in the Au-Au collisions. The solid lines in Figs. 5(a) and 5(b) are the results from Eq. (\ref{59}) with $T_0=0.245$ GeV, which is the only one adjustable parameter. For proton the mass effect was also considered with the transverse kinetic energy $E_T(p_T)=m_T-m_h$, so in fig. 5(b) $p_T$ was replaced by $E_T$. Only the first two components in $\rho^h$ are considered in Fig. 5. To get better fit and widen the $p_T$ and $b$ ranges, the third component generated by TS recombination must be included. 

         \begin{figure*}
        \centering
        \begin{tabular}{ccc}
        \includegraphics[width=0.45\textwidth]{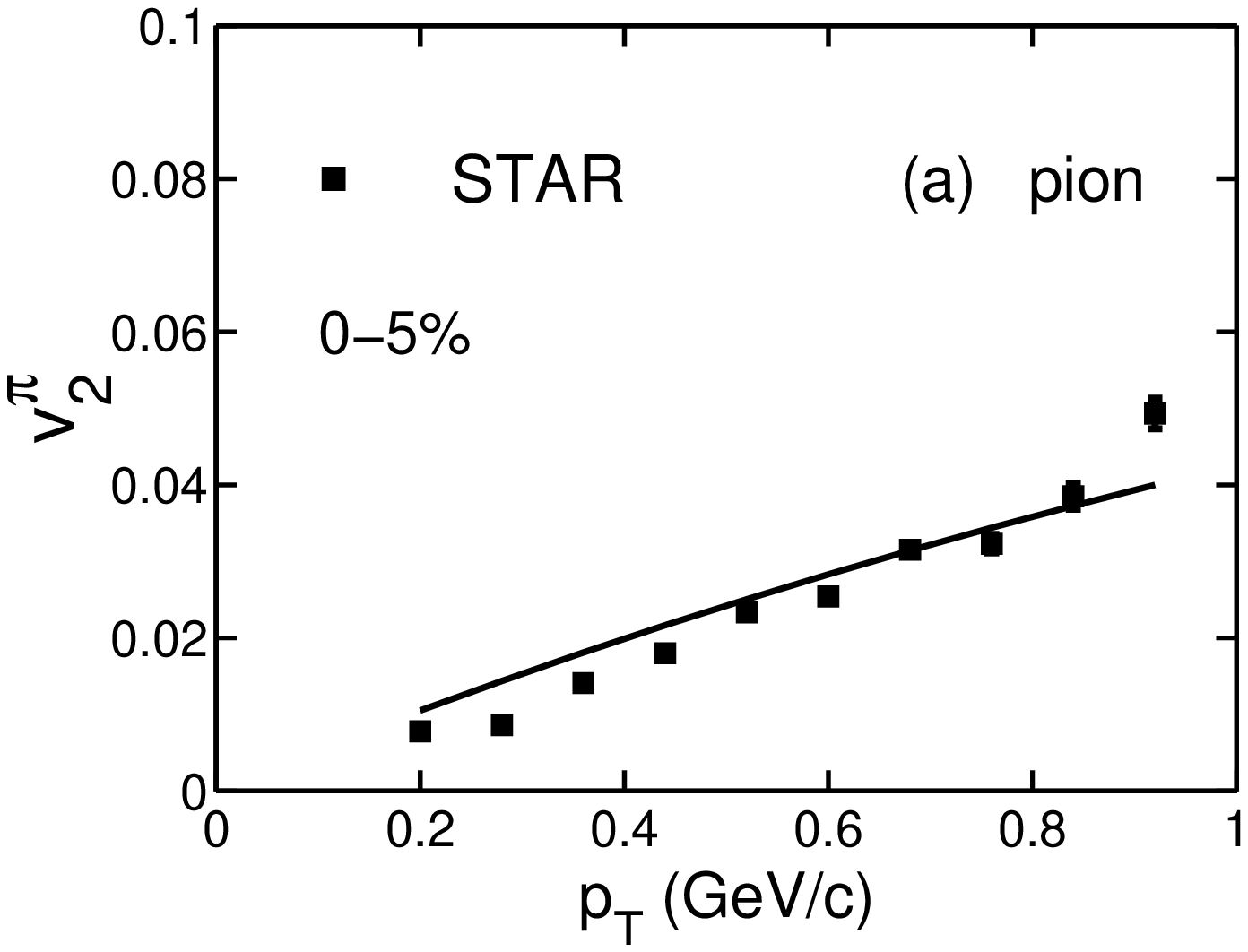}
          \includegraphics[width=0.45\textwidth]{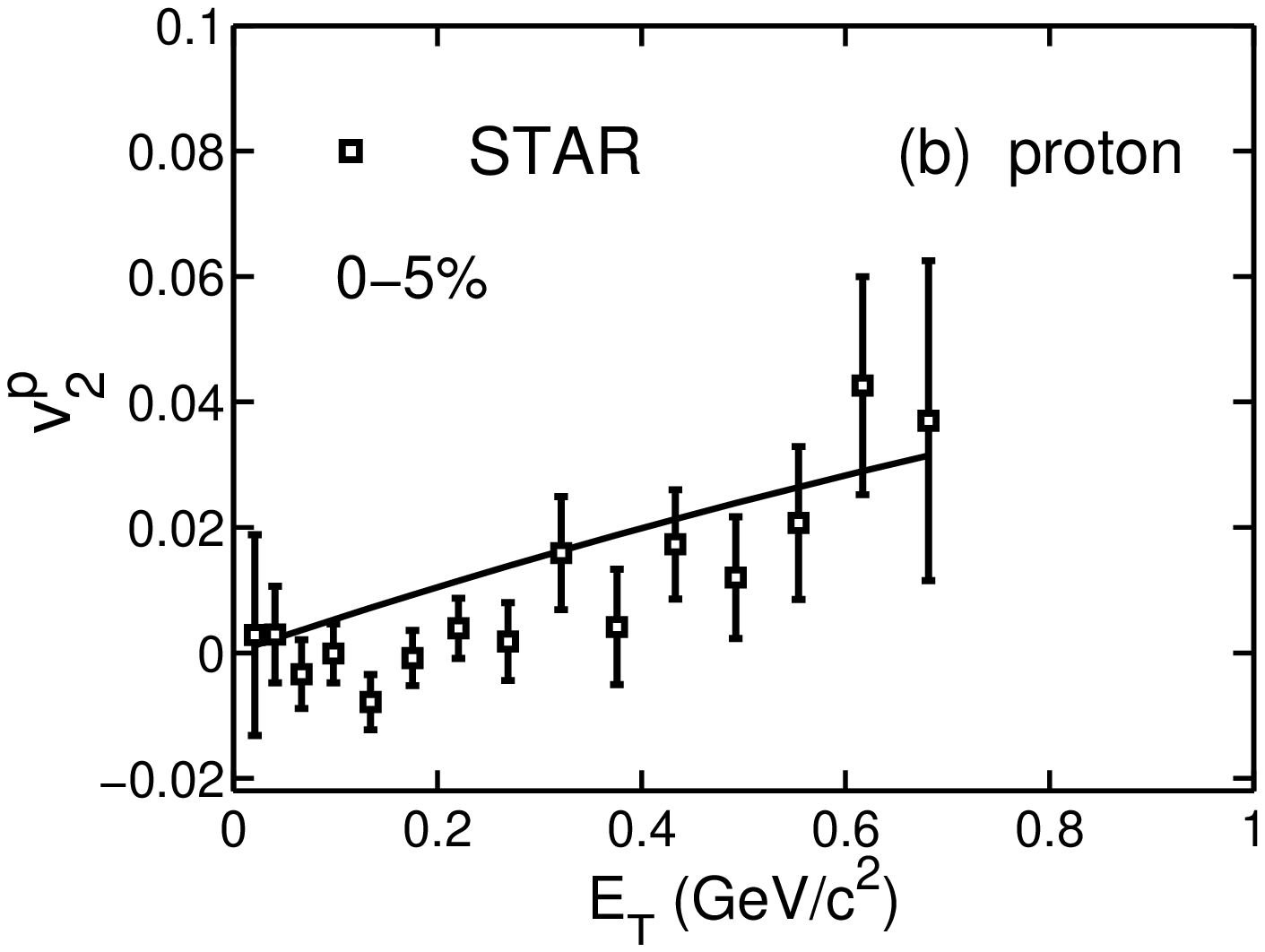}
               \end{tabular}
\caption{(Color online) $v_2$ at 0-5\% centrality for (a) pion and (b) proton. The data are from \cite{STAR2005}.}\label{Fig5}
    \end{figure*}
    
\section{Higher Harmonics}
We have shown that $v_2$ can be understood in terms of the $\phi$ dependence of the TT recombination of the thermal partons affected by the passage of semihard partons through the medium. The space-momentum transformation is accomplished by studying the minijets emitted from the initial elliptical configuration. It is then natural for us to focus on the effects of the same minijets on the higher harmonics. In our formalism the minijets affect the low-$p_T$ region through TS recombination. Since minijets are produced in any given event in unpredictable directions, the average $\phi$ \dis\ can have all terms in a harmonic analysis. The only aspect of the behavior that our formalism has a predictable power is the dependence on $\pt$ and centrality  because the $\phi$-integrated TS component of recombination has already been formulated and parametrized. The third component of $\rho^h(p_T, \phi, b)$ in Eq.\ (\ref{52}) could be written as
\begin{eqnarray}
M^h(p_T, \phi, b)=J( \phi, b) \bar M^h(\pt,b), \label{512}
\end{eqnarray}
where $J( \phi, b)$ describes the $\phi$-dependent part of the minijet contribution, which is assumed to be factorizable from the average $\bar M^h(\pt,b)$ in the same manner as for $R^h(p_T, \phi, b)$ in Eq.\ (\ref{56}). $J( \phi, b)$ is the normalized form of $\tilde J( \phi, b)$
\begin{eqnarray}
J( \phi, b)=\tilde J( \phi, b)\left/ {{1\over 2\pi} \int_0^{2\pi}d\phi \tilde J( \phi, b)}\right.,  \label{513}
\end{eqnarray}
where $\tilde J( \phi, b)$ contains all the harmonic components, $\cos n\phi$, averaged over the tilt angle $\psi_n$, 
\begin{eqnarray}
\tilde J( \phi, b)=1+b \sum_{n=2}^\infty a_n {n\over \pi}\int_{-\pi/2n}^{\pi/2n} d\psi_n \cos n(\phi-\psi_n).  \label{514}
\end{eqnarray}

Including all three components of $\rho^h(p_T, \phi, b)$ in Eq.\ (\ref{52}), we obtain
\begin{eqnarray}
v_n^h(\pt,b)={\left<\cos n\phi\right>_S \bar R^h(\pt,b) + \left< \cos n\phi\right>_J \bar M^h(\pt,b) \over \bar\rho^h(\pt,b)}  ,  \label{515}
\end{eqnarray}
where
\begin{eqnarray}
\bar\rho^h(\pt,b)&=&B^h(\pt,b)+\bar R^h(\pt,b)+\bar M^h(\pt,b),  \label{516} \\
\left< \cos n\phi\right>_J&=&{1\over 2\pi} \int_0^{2\pi} d\phi \cos n\phi J( \phi, b).  \label{517}
\end{eqnarray}
$\left<\cos n\phi\right>_S$ is as defined in Eq.\ (\ref{510}) for any $n$, but it is zero for $n\ge 3$ because of the periodicity of $S( \phi, b)$. Indeed, $\left< \cos n\phi\right>_J$ receives contribution only from the $a_n$ term in Eq.\ (\ref{514}) because of the orthogonality of the harmonics. For non-central collisions we regard $\bar M^{\pi}(\pt,b)$ to be proportional to $C(N_{\rm part})N_{\rm coll}(b)$. It is the normalization of the thermal parton \dis. $N_{\rm coll}(b)$ is the number of binary collisions. We thus  have
\begin{eqnarray}
\bar M^{\pi}(\pt,b)\left.={C(N_{\rm part})N_{\rm coll}(b)\over C(N_{\rm part}^{max})N_{\rm coll}(b=0)}{dN_{\pi}^{\rm TS}\over \pt d\pt}\right|_{b=0} , \label{35}
\end{eqnarray}
It should be pointed out that the decrease of average path length in the medium as the collision becomes more peripheral. Its consequence is that more fraction of the (semi)hard partons can emerge from the medium as $b$ increases. The results are shown in Fig. 6 with the parameters $a_2=0.6$, $a_3=1.6$ and $a_4=1.2$. It's amazing the all calculated curves agree with the data for $p_T$ dependence for the four centralities. Fig. 6(a) is obviously better than fig. 5(a), after the third component in Eq. (\ref{52}) is included. One parameter $a_n$ for each $n$ can affect only the magnitude of $v_n(p_T,b)$, so the excellent reproduction of the $p_T$ and $b$ dependencies reveals the basic attributes of the approach that we have taken to describe the harmonics. The results support our minijet approach to the treatment of azimuthal anisotropy. Minijets are important and can explain all the low-$p_T$ data in the recombination model.

         \begin{figure*}
        \centering
        \begin{tabular}{ccc}
        \includegraphics[width=0.32\textwidth]{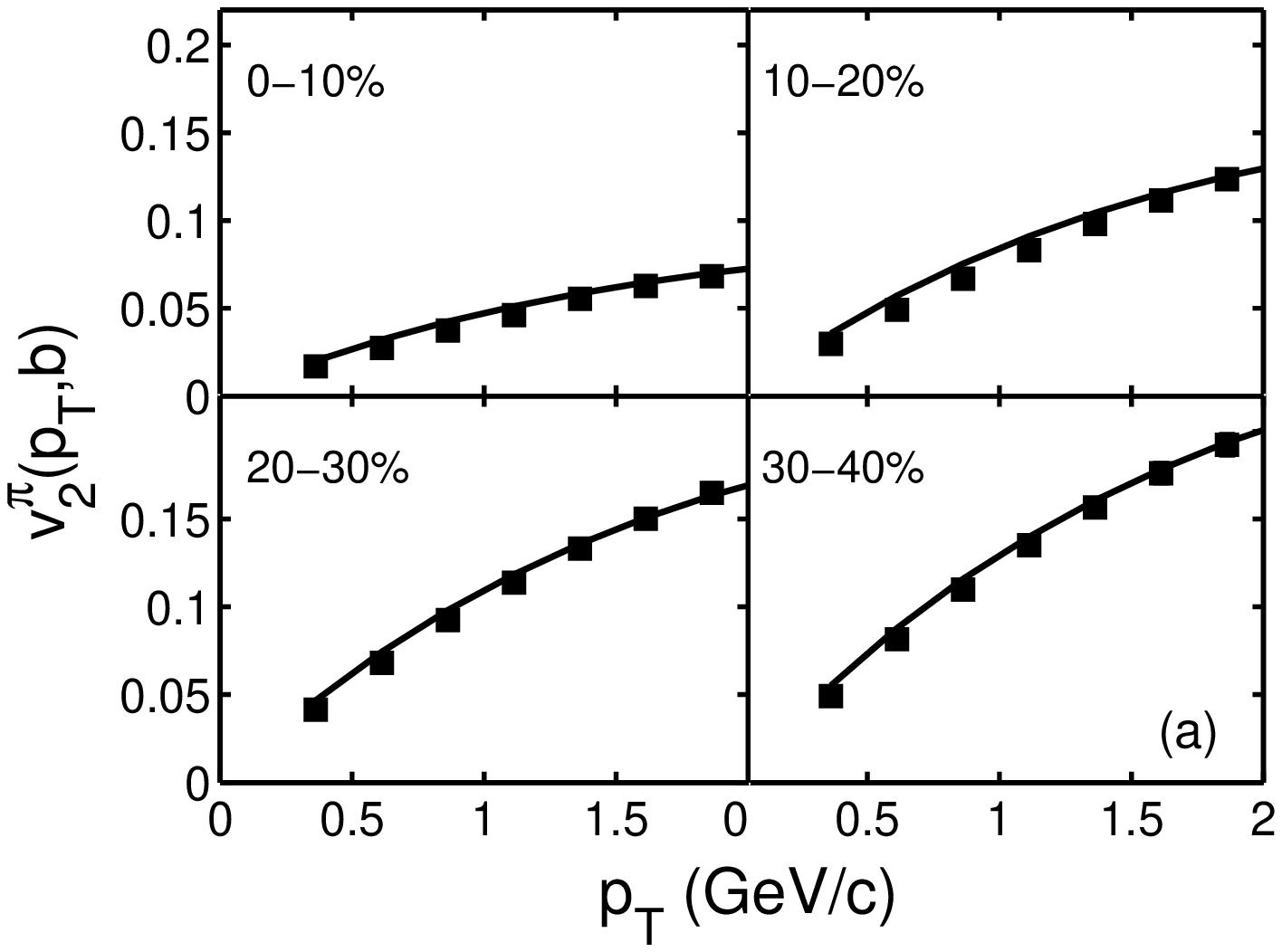}
       \includegraphics[width=0.32\textwidth]{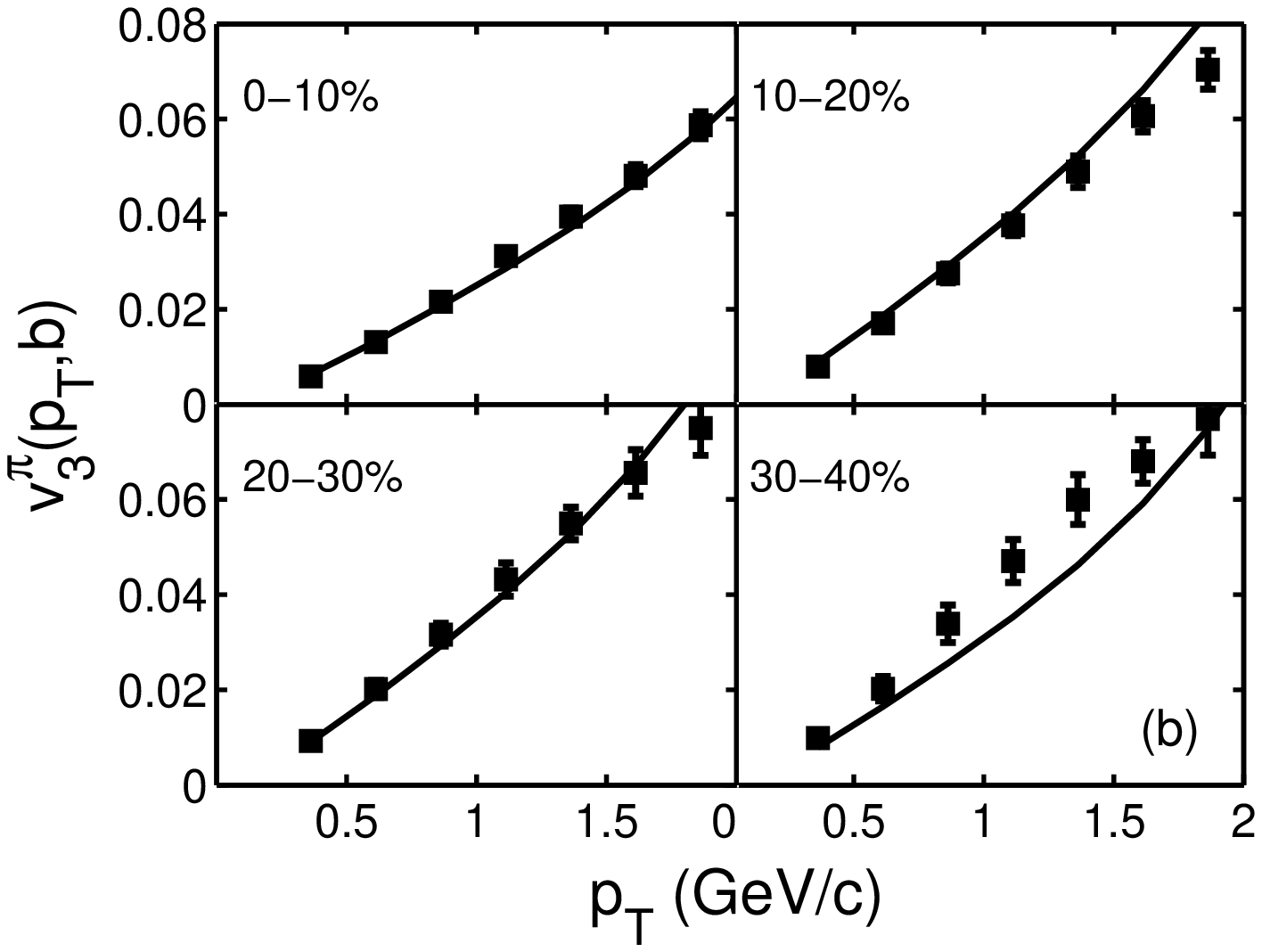}
          \includegraphics[width=0.32\textwidth]{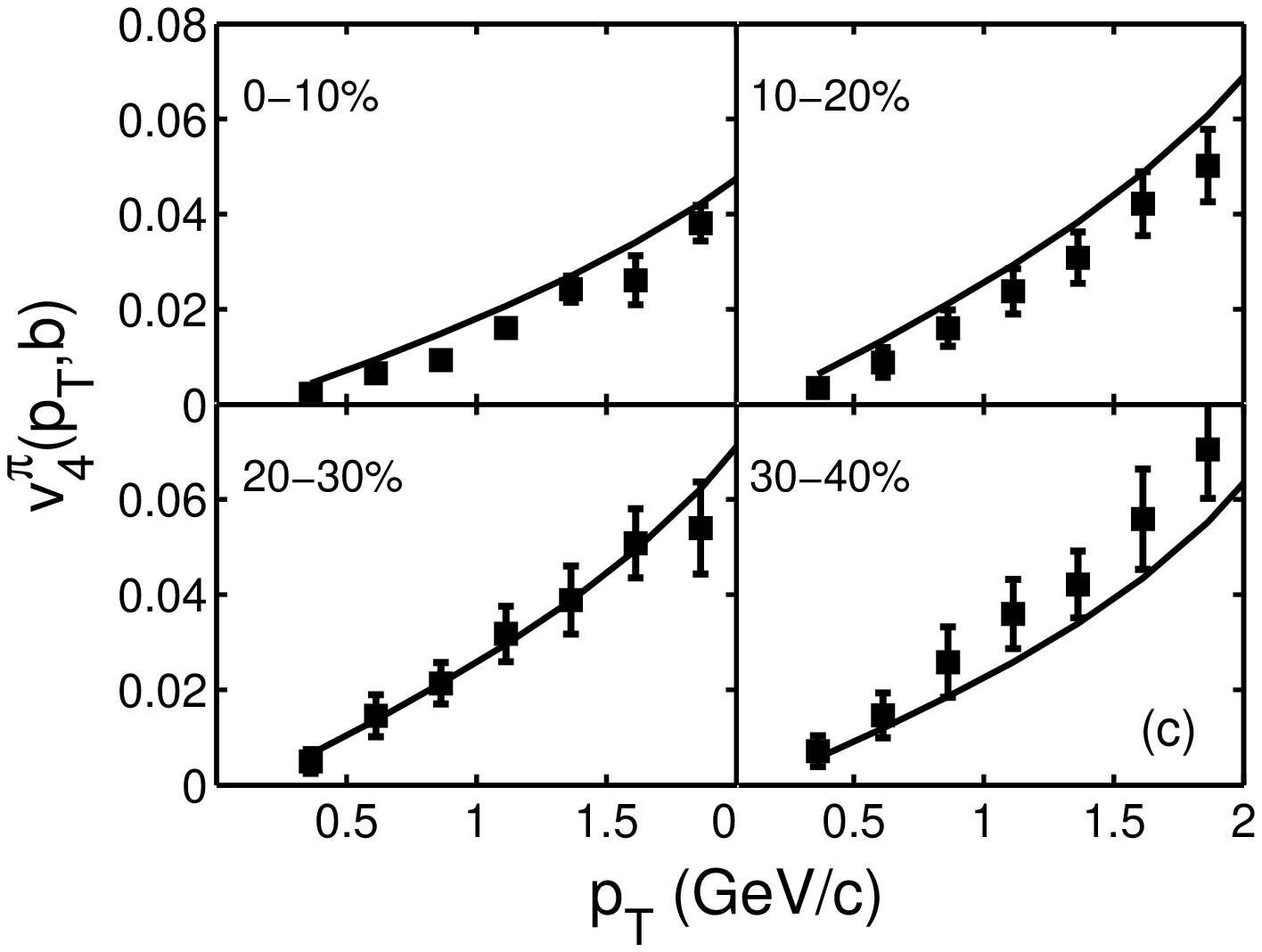}
               \end{tabular}
\caption{(Color online) $p_T$ dependencies of (a) $v_2^\pi(p_T,b)$, (b) $v_3^\pi(p_T,b)$ and (c) $v_4^\pi(p_T,b)$  for four centralities in each case. Data are from Ref.\ \cite{PHENIX2011}.}\label{Fig6}
    \end{figure*}

\section{Summary and outlook}
In this review, we have presented the production of the identified hadrons in Au-Au collisions at RHIC and in Pb-Pb collisions at LHC in a formalism that displays all the components of thermal- and shower-parton recombination. It established that any theoretical treatment of hadrons produced at low and intermediate $p_T$  region would be incomplete without taking the effects of minijets into account. Minijets are important and can explain all the low transverse momentum data in the recombination framework.

For Au-Au collisions at 200 GeV, we have shown that the hadron spectra and azimuthal harmonics can be obtained by taking into account the azimuthal dependence of minijet.  But for Pb-Pb collisions at 2.76 TeV we have only investigated the case of central collisions. The consideration for the central collisions only represents the first, but significant, step toward understanding the physics of hadronization at LHC. To extend the study to non-central collisions at LHC is the natural problem to pursue next. How minijets influence the azimuthal harmonics measured at LHC will be a major area of investigation. 

\section*{Conflict of Interests}
The authors declare that there is no conflict of interests regarding the publication of this article.

\section*{Acknowledgments}
This work is supported by the NSFC of China under Grant No.\ 11205106.

\end{document}